\documentclass[final,5p,twocolumn]{elsarticle}
\usepackage[english]{babel}
\usepackage{caption}
\usepackage{fontenc}
\usepackage{subfigure}
\usepackage{amsmath}
\usepackage{subeqnarray}
\usepackage[all]{xy}
\usepackage{ctable}
\usepackage{stfloats}
\usepackage{dcolumn}
\usepackage{array}
\usepackage{amssymb}
\usepackage[version=3]{mhchem}
\usepackage{bbding}
\usepackage{longtable}
\usepackage{ifsym}
\usepackage{natbib}
\biboptions{sort&compress}
\usepackage[skip=-1pt]{caption}
\usepackage{titlesec}
\titlespacing*{\section}
{0pt}{1.5ex plus 1ex minus .2ex}{1ex plus .2ex}
\titlespacing*{\subsection}
{0pt}{1.5ex plus 1ex minus .2ex}{1ex plus .2ex}

\graphicspath{{./img/}}

\usepackage{graphicx,color,psfrag}

\newcommand{\p}{\partial}

\catcode`\@=11
\def\gtsim{\mathrel{\vcenter{\m@th\offinterlineskip
\hbox{$\hfill>\hfill$}\kern.5ex\hbox{$\hfill\sim\hfill$}}}}
\catcode`\@=12

\catcode`\@=11
\def\ltsim{\mathrel{\vcenter{\m@th\offinterlineskip
\hbox{$\hfill<\hfill$}\kern.5ex\hbox{$\hfill\sim\hfill$}}}}
\catcode`\@=12
\catcode`\@=12

\begin{document}
\baselineskip 18pt
\title{Regimes of boundary-layer ignition by heat release from a localized energy source}


\author[uc3m]{Mario S\'anchez-Sanz}
\author[uc3m]{Eduardo Fern\'andez-Tarrazo}
\author[ucsd]{Antonio L. S\'anchez}

\journal{Combustion and Flame}

\address[uc3m]{Dept. Ingenier\'{\i}a T\'ermica y de Fluidos, Universidad Carlos 
III de Madrid, Legan\'es 28911, Spain}
\address[ucsd]{Dept. Mechanical and Aerospace Engineering, University of 
California San Diego, La Jolla CA 92093-0411, USA}

\begin{abstract}

This paper investigates the initiation of a deflagration in a premixed boundary-layer stream by continuous heat deposition from a line energy source placed perpendicular to the flow on the wall surface, a planar flow configuration relevant for small-scale combustion applications, including portable rotary engines. Ignition is investigated in the constant density approximation with a one-step irreversible reaction with large activation energy adopted for the chemistry description. The ratio of the characteristic strain time, given by the inverse of the wall velocity gradient, to the characteristic deflagration residence time defines the relevant controlling Damk\"ohler number $D$. The time-dependent evolution following the activation of the heat source is obtained by numerical integration of the energy and fuel conservation equations. For sufficiently small values of $D$, the solution evolves towards a steady flow in which the chemical reaction remains confined to a finite near-source reactive kernel. This becomes increasingly slender for increasing values of $D$, corresponding to smaller near-wall velocities, until a critical value $D_{c1}$ is reached at which the confined kernel is replaced by a steady anchored deflagration, assisted by the source heating rate, which develops indefinitely downstream. As the boundary-layer velocity gradient is further decreased, a second critical Damk\"ohler number $D_{c2}>D_{c1}$ is reached at which the energy deposition results in a flashback deflagration propagating upstream against the incoming flow along the base of the boundary layer. The computations investigate the dependence of $D_{c1}$ and $D_{c2}$ on the fuel diffusivity and the dependence of $D_{c1}$ on the source heating rate, delineating the boundaries that define the relevant regime diagram for these combustion systems.

\end{abstract}


\maketitle

\section{Introduction}

Near-wall combustion processes are critical for the successful operation of many reactive systems. For instance, Lean-Premixed burners are prone to flashback, or upstream propagation of the flame in the premixing tubes, which may compromise the combustor integrity. The first description of laminar-flame propagation near a wall under the effect of a velocity gradient was given by Lewis and von Elbe \cite{Lewis1943}. They found that flame flashback occurs if the gas velocity gradient at the wall $A$ is below a critical value $A_c$ of the order of the ratio of the laminar flame speed $S_L$ and the quenching distance, the latter scaling with the flame thickness $\alpha_{\scriptscriptstyle T}/S_L$, where $\alpha_{\scriptscriptstyle T}$ is the characteristic value of the thermal diffusivity of the gas mixture.  
This critical value $A_c \sim S_L^2/\alpha_{\scriptscriptstyle T}$ can be determined numerically, as done by Kurdyumov et al. \cite{KFL_2000} and Kurdyumov \& Li\~n\'an \cite{KL_2002}, who considered the propagation of a premixed flame along the near-wall, low-velocity region found at the base of a laminar boundary layer. In agreement with the Lewis and von Elbe criterion, they found that the relevant controlling parameter is the Damk\"ohler number  $D=S_L^2/(\alpha_{\scriptscriptstyle T} A)$, so that the flame propagates against the incoming flow only for values of the Damk\"ohler number  $D$ larger than a critical value of order unity. Gruber et al. \citep{gruber2012direct,gruber2015modeling} tackled the problem of flashback in turbulent flames. In their work, they studied numerically the transient upstream propagation of hydrogen premixed flames and found that the interaction of the Darrenius-Landau instability with the turbulence was to blame for the onset of flashback.
Relevant experimental  studies of the flashback phenomenon include that of Eichler and Sattelmayer \cite{Eichler2012} for methane and methane-hydrogen blends. In their experiments, the fuel-air stream was continuously ignited by two pilot burners, providing a steady boundary-layer flow in the absence of flashback. Also relevant to the present study is the work of Ishida \cite{Ishida_2011}, who studied the effect of the airflow velocity on flame spread over a fuel-soaked surface, including the determination of the critical conditions that lead to flame blow off and flame flashback.

Near-wall flame propagation is particularly relevant in connection with small combustion devices, such as rotary engines, which are subject to short residence times, elevated heat losses, and high strain rates \cite{DR_2005}. The performance of these systems relies on the rapid propagation of the premixed flame generated at the wall by a glow plug \cite{Sprague_2007,Tsuji_2005}. Most previous studies of ignition by an external energy source, concerned with the determination of the minimum energy needed to produce a successful deflagration (e.g. \cite{Warnatz1988,KurdyumovMIE2003}), consider a stagnant mixture or study the transient ignition of boundary layers developing over heated flat surfaces \citep{lee1986transient,law1981flat}. Interactions of the external source with the flow strain are fundamental for flame initiation in boundary-layer flows, a problem to be considered here.

\begin{figure}[htp]
 \centering
  \includegraphics[width=.27\textwidth,angle=0]{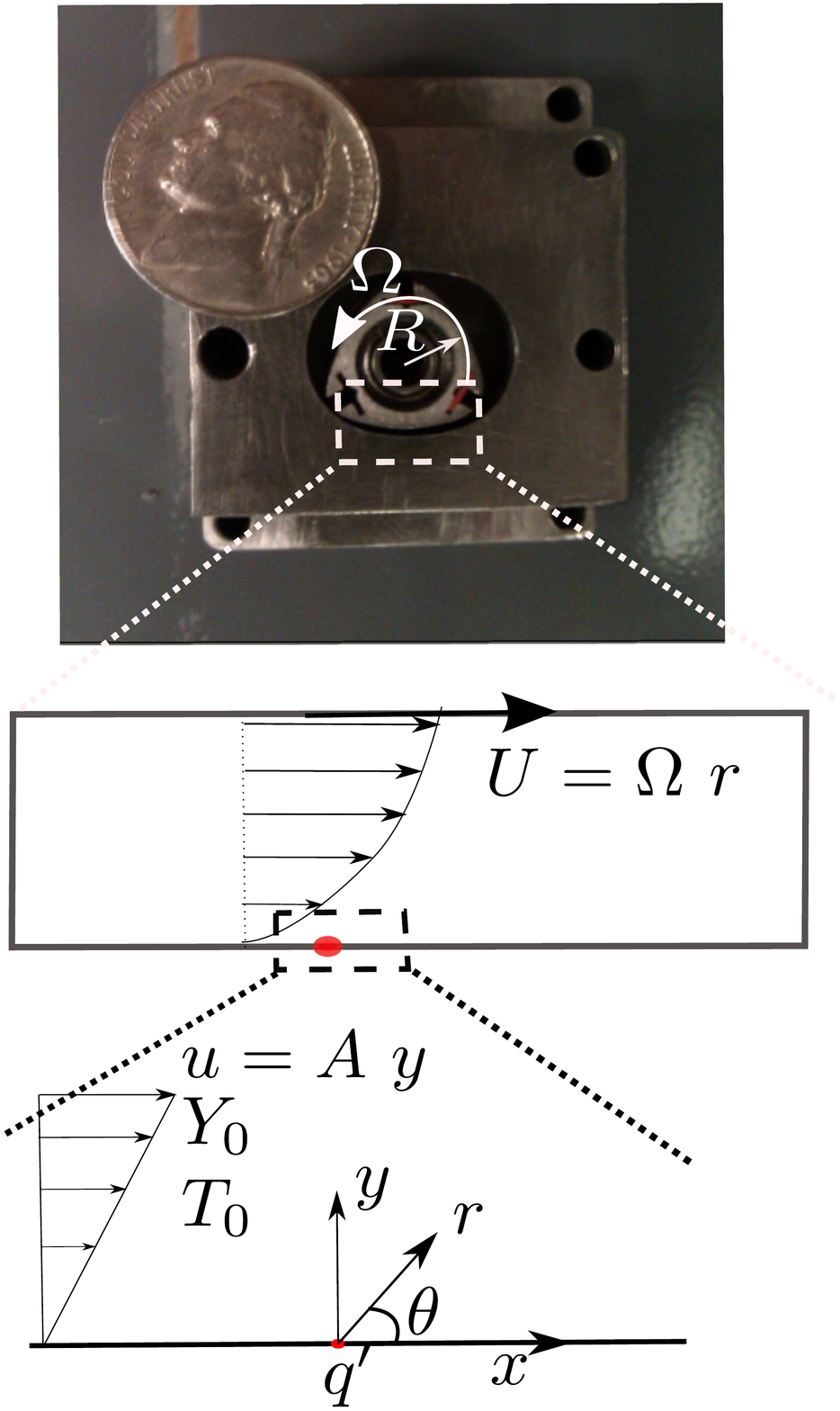}
  \caption{A schematic representation of the flow. The picture illustrates the rotary engine developed by the group of C. Fern\'andez-Pello in the University of California in Berkeley \cite{Sprague_2007}}
\label{fig:sketch}
\end{figure}

The present paper addresses combustion phenomenology that is relevant to small-scale rotary engines, as that shown in Fig. \ref{fig:sketch}. The engine includes a triangular rotor that revolves inside the epitrochoidal housing. The rotor admits a combustible mixture that is mechanically compressed to reach a maximum pressure when the volume of the gas confined between the rotor and the housing is minimum. At that precise instant the mixture is ignited, generating a deflagration that propagates in the compressed gas away from the ignition point. The geometry of the combustion chamber, which changes continuously as a result of the relative motion of the confining walls, is very slender, its characteristic length being about twenty times its width, and the latter being of the order of the flame thickness \cite{Sprague_2007,Tsuji_2005}. The final efficiency of the system can be quantified as the fraction of the chemical energy of the fuel that is transformed into mechanical energy. 

To promote fuel flexibility and improve performance, the combustible mixture is typically ignited using a glow plug. The power output is maximized using large rotational velocities in the range of 3,000 to 20,000 rpm \cite{Sprague_2007}. With such rotational speeds, the ignition of the fuel mixture takes place in a gas subjected to large velocity gradients, on the order of $10^3$ s$^{-1}$ for the rotary engine of Fig.~\ref{fig:sketch}. While the high temperatures found in the vicinity of the glow plug ensure the initiation of the chemical reaction there, successful flame propagation from this near-source reactive kernel depends upon the interplay of the chemical reactions with the heat and mass transfer processes occurring in the near-wall strained flow, with relevant governing parameters including the intensity of the energy source, the wall velocity gradient, and the mass diffusivity of the fuel, among others. The associated phenomenology, including the different combustion regimes, is to be investigated below with use made of a model problem that provides the needed understanding.

\section{Formulation}

To help clarify the complex interactions between the igniter and the near-wall flow in micro rotary engines we investigate here unsteady deflagration-initiation processes induced by the sudden application of a localized energy source of constant heating power $q'$. As shown in the sketch of Fig.~\ref{fig:sketch}, we consider a lean reactive mixture with fuel mass fraction $Y_0$ and temperature $T_0$ flowing parallel to a flat adiabatic wall. To attain maximum simplification, the density and transport properties are assumed to be constant in this initial study, so that the velocity field in the near-wall region is given at all times by $u=A y$ in terms of the distance to the wall $y$. The chemistry is modeled with an irreversible reaction with rate (mass of fuel consumed per unit volume and unit time) given in terms of the local fuel mass fraction $Y$ and temperature $T$ by the Arrhenius expression $\omega=\rho Y B \exp(-T_a/T)$, involving the gas density $\rho$, the frequency factor $B$, and the activation temperature $T_a$. The associated Zeldo'vich number $\beta=(T_e-T_0)T_a/T^2_e$ is taken to be large in our analysis, resulting in an exponentially small rate of reaction in the absence of heat deposition, with $\beta=10$ selected in the computations shown below. Here, $T_e=T_0+Q Y_0/c_p$ is the adiabatic flame temperature, which is expressed in terms of the amount of heat released per unit mass of fuel burnt $Q$ and the specific heat at constant pressure $c_p$. 

\subsection{The mathematical problem}

The description employes polar coordinates $r$ and $\theta$ centered at the energy source, with the radial distance $r$ scaled with the characteristic size $\delta_c=(\alpha_{\scriptscriptstyle T}/A)^{1/2}$ of the hot kernel affected by the heat released from the energy source, determined by a convection-diffusion balance in the near-wall region. The characteristic velocity in this hot kernel $v_c=A \delta_c=(\alpha_{\scriptscriptstyle T} A)^{1/2}$ and the associated residence time $\delta_c/v_c=A^{-1}$ are used to define a dimensionless flow velocity $(v_r,v_\theta)=(r \cos \theta \sin \theta,-r \sin^2 \theta)$ and a dimensionless time $t$. The temperature and the fuel mass fraction are expressed in the normalized form $\hat{T}=(T-T_0)/(T_e-T_0)$ and $\hat{Y}=Y/Y_0$. Writing the energy and fuel conservation equations in terms of these dimensionless variables yields
\begin{align}
\frac{\p \hat{T}}{\p t} & + r \cos \theta \sin \theta \frac{\p \hat{T}}{\p  r}-\sin^2 \theta \frac{\p  \hat{T}}{\p \theta}= \nonumber \\ 
&\frac{1}{r} \frac{\p}{\p r} \left(r  \frac{\p \hat{T}}{\p  r}\right) +\frac{1}{r^2} \frac{\p^2 \hat{T}}{\p \theta^2}+\Omega \label{ener1} \\ 
\frac{\p \hat{Y}}{\p t} & + r \cos \theta \sin \theta \frac{\p \hat{Y}}{\p  r}-\sin^2 \theta \frac{\p  \hat{Y}}{\p \theta}= \nonumber \\ 
&\frac{1}{L_F} \left[\frac{1}{r} \frac{\p}{\p r} \left(r \frac{\p \hat{Y}}{\p r}\right) +\frac{1}{r^2} \frac{\p^2 \hat{Y}}{\p \theta^2}\right]-\Omega 
\label{Y1} 
\end{align}
where $L_F$ is the fuel Lewis number and $\Omega=\omega/(\rho Y_0 A)=\beta^2/(2 L_F) D \hat{Y} \exp[\beta(\hat{T}-1)]$
is the dimensionless reaction rate, involving the Damk\"ohler number
$D=S_L^2/(\alpha_{\scriptscriptstyle T} A)$.
For convenience in expressing the reaction rate, the frequency factor $B$ of the Arrhenius law has been related to the steady propagation velocity of the planar deflagration $S_L$ with use made of $S_L=(2 L_F \alpha_{\scriptscriptstyle T} \beta^{-2} B e^{-T_a/T_e})^{1/2}$, obtained at leading order in the asymptotic limit $\beta \gg 1$. The latter limit has been also considered when writing the temperature dependence of the reaction rate $\Omega$ in the simplified form $\exp\{[\beta(\hat{T}-1)]/[1+(T_e-T_0)(\hat{T}-1)/T_e] \simeq \exp[\beta(\hat{T}-1)]$.
The above equations must be integrated with initial conditions $\hat{T}=\hat{Y}-1=0 \quad {\rm at}$  at  $t=0$
as corresponds to the initially unperturbed cold mixture. The boundary conditions for $t>0$ are
$\p \hat{T}/\p \theta=\p \hat{Y}/\p \theta=0$ at  $\theta=0$ and $ \theta=\pi$ for $r>0$,
corresponding to a noncatalytic adiabatic wall, and
$\hat{T}=\hat{Y}-1=0$ for $r \rightarrow \infty$ and $\pi \ge \theta\ge0$,
stating that the flow remains unperturbed in the far field. On the other hand, the boundary condition corresponding to a line heat source located at $r=0$ is
\begin{equation}
r \frac{\p \hat{T}}{\p r}+ q= \hat{Y}=0 \quad {\rm as} \quad r \rightarrow 0 \quad {\rm for} \quad 0\le \theta \le \pi, \label{bc2}
\end{equation}
where $q=q'/(\pi \kappa (T_e-T_0))$ represents an appropriately scaled measure of the heat deposition rate, with $\kappa$ denoting the thermal conductivity. The associated near-source  temperature distribution is given by $\hat{T}=C(t)-q \ln r$. The logarithmic divergence is consistent with the chemical-equilibrium condition $\hat{Y}=0$ as $r \rightarrow 0$ used in~\eqref{bc2}.


\subsection{Numerical method}

The problem (\ref{ener1})-(\ref{Y1}), with the corresponding inital and boundary conditions stated above, was integrated by marching in time with a semi-implicit finite-difference scheme, second order in both time and space. Once the solution is known at time $t$, both the spatial derivatives and the reaction terms are discretized at $t + \delta t$. The resulting non-linear system of equations is solved using an iterative procedure. The iteration $k$ is initiated by solving the mass fraction equation, with the exponential of the reaction term evaluated at time $t$. Using the new value of the mass fraction $\hat{Y}^{k+1}$, we solve the energy equation to obtain the new value of the temperature $\hat{T}^{k+1}$, again using in the exponential of the reaction term the temperature of the previous iteration $\hat{T}^{k}$. The procedure continues until the difference between two consecutive iterations $|\hat{T}^{k}-\hat{T}^{k+1}| + |\hat{Y}^{k}-\hat{Y}^{k+1}|$ is below a given tolerance $\epsilon$, with $\epsilon=10^{-4}$ used in most computations. The number of iterations needed to achieve convergence depends on the specific values of the parameters and on the time step $\delta t$ used in the integration. Typically, no more than 25-30 iterations were needed for $\delta t=0.05$, the time step selected for the computations discussed below.

Because of the singular behavior of the temperature as $r \to 0$, the boundary condition~\eqref{bc2} was implemented at a small distance from the source $r=R_{min} \ll 1$. The temperature at the first point of the grid $j=1$ was computed according to $\hat{T}_{j=1}=\hat{T}_{j=2}-q \ln \left( R_{min}/r_{j=2} \right)$, consistent with the logarithmic behavior previously identified. The integrations employed a uniform grid in the azimuthal direction with $\delta \theta = \pi/150$ and a non-uniform grid in the radial direction, with minimum spacing near $r=R_{min}$ increasing towards the outer edge of the computational domain $r = R_{max}$. The values of $R_{min}$, $R_{max}$ and the maximum and minimum radial grid spacing were varied depending on the conditions, so that, for instance, a smaller grid spacing was needed to provide accurate results when computing flame initiation  events induced by higher heating rates $q$. Representative values are $R_{min}=10^{-3}$, $R_{max}=100$, $\delta r_{min}=0.02$, and $\delta r_{max}=0.14$.

\section{Ignition regimes}

The evolution of the reactive flow after the source is activated at $t=0$ is determined by integration of~\eqref{ener1} and~\eqref{Y1} with the initial and boundary conditions, indicated above. As discussed below, three different reaction regimes are obtained depending on the values of the controlling parameters, namely, a confined reactive kernel surrounding the heat source, an anchored deflagration that extends downstream, and an upstream propagating deflagration. For given values of $\beta$,  $q$, and $Le$ the transition between regimes occurs at critical values of the Damk\"ohler number $D$. Transition from a confined reactive kernel to an anchored deflagration occurs when the residence time in the near-source region $A^{-1}$ exceeds a critical value of the order of the characteristic chemical-heat-release time $\alpha_{\scriptscriptstyle T}/S_L^2$, thereby defining a critical Damk\"ohler number $D_{c1}$. On the other hand, transition from an anchored deflagration to an upstream  propagating deflagration occurs when the near-wall flow velocity $(\alpha_{\scriptscriptstyle T} A)^{1/2}$ decreases below the propagation velocity of the curved deflagration, a competition that is measured by a second critical Damk\"ohler number $D_{c2}$.

\subsection{Confined reactive kernel}
\begin{figure}[ht]
 \centering
  \includegraphics[width=.375\textwidth]{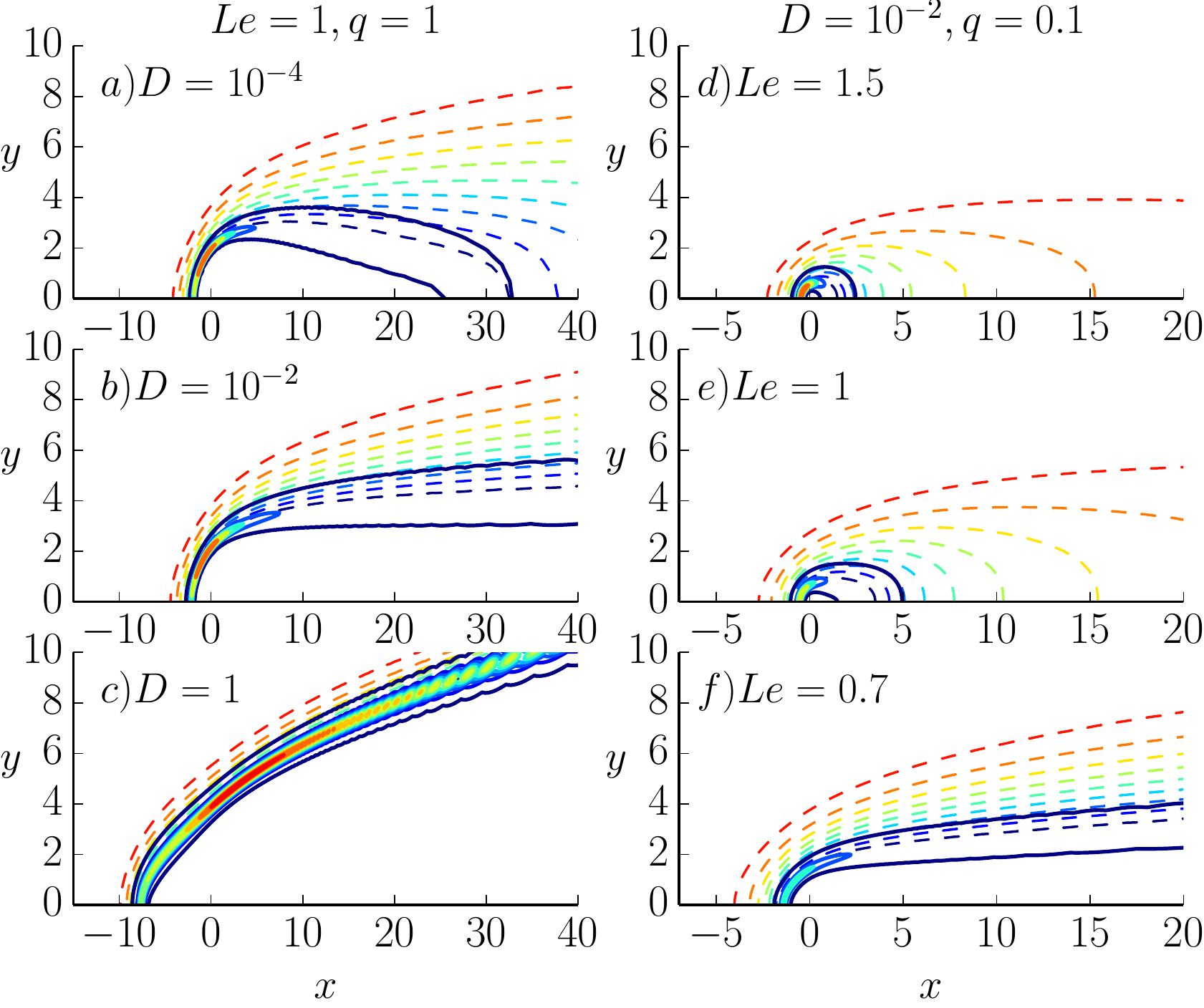}
  \caption{Sample near-source solutions obtained for $t \gg 1$ by integration of~\eqref{ener1} and~\eqref{Y1} with different values of $D$, $q$, and $Le$, as indicated in the labels. Each plot includes isocontours of reaction rate $\Omega$ (solid curves) and mass fraction $\hat{Y}$ (dashed curves) with equal increments $\delta \Omega=\Omega_{max}/10$ and $\delta \hat{Y}=0.1$, respectively.}
\label{fig:sample}
\end{figure}

For values of the Damk\"ohler number below a critical value $D_{c1}(q,Le)$ the solution evolves towards a steady flow in which the chemical reaction is confined to a reactive kernel surrounding the heat source, where the fuel is depleted. Sample steady solutions of this type are shown in Figs.~\ref{fig:sample}(a), \ref{fig:sample}(d), and~\ref{fig:sample}(e). In this case, because of the low reactivity of the mixture, the chemical reaction depends entirely on the heat provided by the energy source, flame propagation being precluded by the existing large velocity gradients.  
Because of the limited extent of the chemical reaction, the perturbations to the far field resulting from the presence of the source are necessarily confined to a near-wall wake region extending for $\theta \sim r^{-2/3}\ll 1$ when $r \gg 1$, where we find small perturbations $\hat{T} \sim 1-\hat{Y} \sim r^{-2/3} \ll 1$. Correspondingly, the associated self-similar solution can be described in terms of the rescaled coordinate $\eta=\theta r^{2/3}$ and rescaled variables  $\mathcal{T}=r^{2/3}\hat{T}$ and  $\mathcal{Y}=(1-\hat{Y})r^{2/3}$, which reduce (\ref{ener1}) and (\ref{Y1}) to $\tfrac{2}{3}\eta^3 \left(\mathcal{T}/\eta \right)' - \mathcal{T}' \eta^2 = \mathcal{T}'' $ and $\tfrac{2}{3}\eta^3 \left(\mathcal{Y}/\eta \right)' - \mathcal{Y}' \eta^2 = \mathcal{Y}''/Le $, respectively, with the prime $'$ denoting differentiation with respect to $\eta$. Integrating these equations with boundary conditions $\mathcal{T}=\mathcal{Y}=0$ as $\eta \to \infty$ and $\mathcal{T}'=\mathcal{Y}'=0$ at $\eta =0$ yields 
$\mathcal{T}=\mathcal{T}_w \exp[-\eta^3/9]$  and $ 
\mathcal{Y}=\mathcal{Y}_w \exp[-\eta^3 L_F/9].$
The constant factors $\mathcal{T}_w$ and $\mathcal{Y}_w$ are to be determined numerically, giving , for instance, $\mathcal{T}_w=2.292$ and  $\mathcal{Y}_w=1.805$ for $q=0.1, D=0.01, Le=1$. 

\subsection{Anchored deflagration}

As expected, the kernel becomes larger for increasing values of $D$, corresponding to more reactive mixtures or smaller velocity gradients, and also for increasing heating rates $q$. For the same Damk\"ohler number, a higher fuel diffusivity (i.e., smaller values of $Le$) also favors the growth of the reactive kernel, as is evident from the comparison of the results shown in Figs.~\ref{fig:sample}(d) and~\ref{fig:sample}(e). Because of the effect of convection, the growth of the reactive region is predominantly in the streamwise direction, yielding an increasingly slender kernel. 

The confined steady kernel can no longer exist for values of the Damk\"ohler number $D > D_{c1}$. Instead, the solution that emerges for large times includes a curved deflagration that originates in the near source reactive kernel, developing downstream for increasing times. Although the flow continues to evolve in time indefinitely in the far-field downstream region, the temperature and fuel mass fraction at distances $r$ of order unity approach for large times steady distributions corresponding to an anchored deflagration assisted by the heating rate. Figures~\ref{fig:sample}(b), \ref{fig:sample}(c), and~\ref{fig:sample}(f) correspond to sample computations of these anchored-flame solutions. 

The heat coming from the source is essential for the existence of these anchored flames, in that the flame would be readily blown off downstream should the source be turned off. The influence of the source diminishes for increasing distances downstream from the source and becomes entirely negligible in the far field, where the shape of the resulting curved reactive front $y_f \propto \sqrt{x}$ is determined by the balance between the propagation velocity of the deflagration and the local flow velocity. 

\subsection{Upstream propagating deflagration}

The anchored flame becomes more robust as the Damk\"ohler number increases from $D_{c1}$. With increased reactivity, the front becomes less dependent on the heat addition from the source and can correspondingly migrate farther upstream, a change that is apparent when comparing the results in Figs.~\ref{fig:sample}(b) and \ref{fig:sample}(c). Anchored flames cease to exist for values of Damk\"ohler number exceeding a critical value $D_{c2}(Le)$, above which the heat deposition from the source leads instead to the formation of a flame front that propagates indefinitely upstream, corresponding to the flashback mode investigated earlier \cite{KFL_2000,KL_2002}. For $Le=1$ the critical Damk\"ohler number for flashback is $D_{c2}=1.07$, suggesting that the case $D=1$ depicted in Fig.~\ref{fig:sample}(c) corresponds to near-marginal conditions, the resulting flame shape being necessarily similar to that of the propagating flame found for $D=1.07$.

\begin{figure}[ht]
 \centering
  \includegraphics[width=0.45\textwidth]{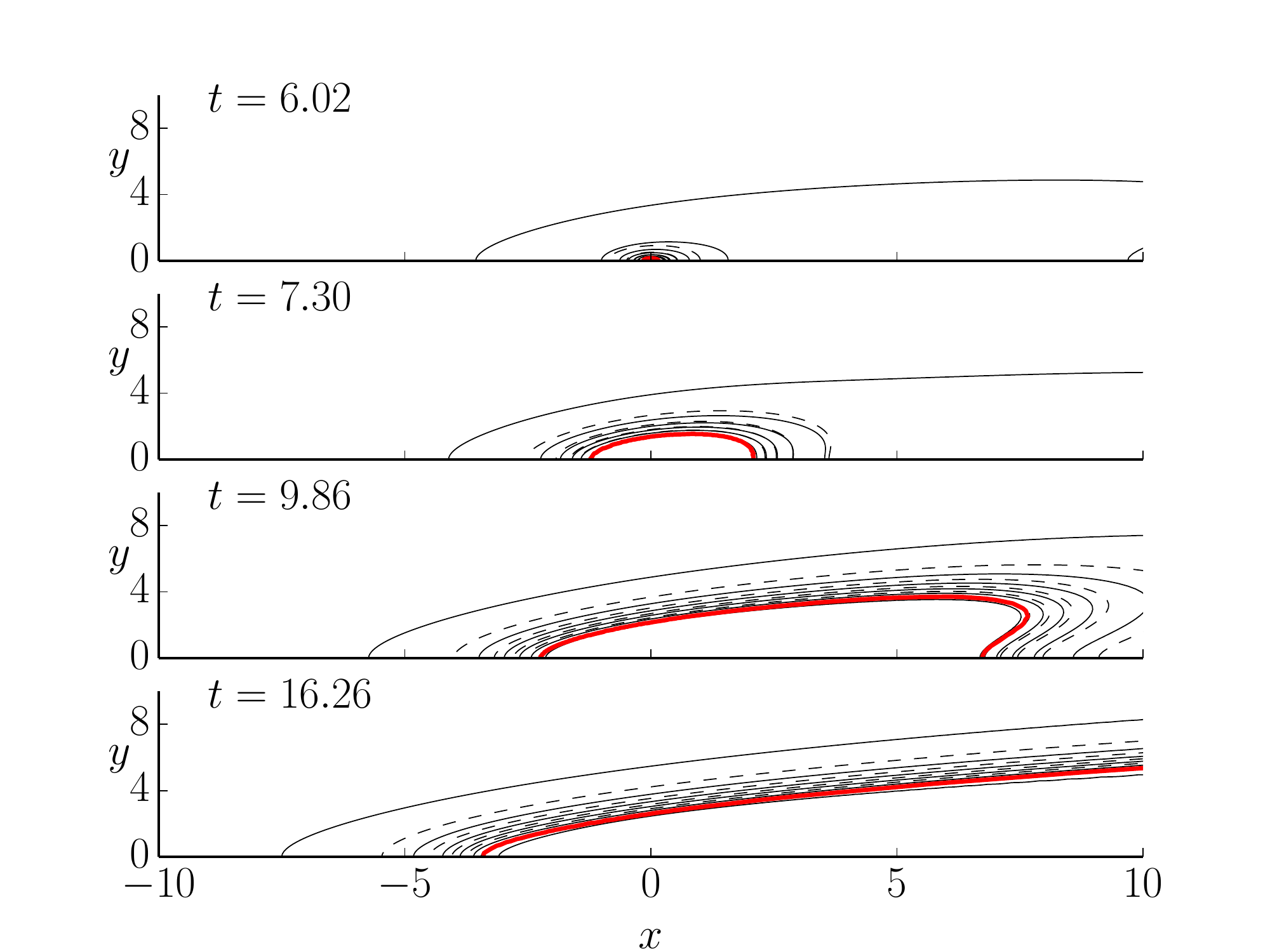}
  \caption{Flame location $r_{max}(\theta,t)$ (thick curve) and associated isocontours of temperature (thin-solid curves at incremets of $\Delta T=0.2$ from $T=0$) and fuel mass fraction (dashed curves at incremets of $\Delta Y=0.2$ from $Y=0.1$) obtained for $q=0.1$, $Le=1$, and $D=1.1>D_{c2}$ at different instants of time.}
\label{fig:profilesD>Dc2}
\end{figure}

The typical time evolution of a flame with $Le=1$  is shown in Fig.~\ref{fig:profilesD>Dc2} for a marginally supercritical case with $D=1.1>D_{c2}=1.07$. The flame location is identified at each time $t$ by the radial location $r_{max}(\theta,t)$ where $\Omega$ reaches a maximum for each value of $\theta$. Besides $r_{max}(\theta,t)$, the plot shows the isocontours of fuel and temperature at different instants of time following the activation of the energy source. The propagating deflagration, which bounds a central region depleted from fuel, moves away from the source, with a downstream velocity that is considerably faster as a result of the convective flow. 
Initially the reaction layer receives heat from the source and the resulting motion is very rapid, giving a rapid transition between $t=6.02$ and $t=7.30$. The upstream motion slows down progressively as the  temperature gradient behind the reaction layer diminishes. Eventually, the flame front moves independently of the source, leaving the gas behind at uniform temperature, equal in this case $Le=1$ to the adiabatic flame temperature $T=1$. For this near marginal case the resulting propagation velocity is very slow.

\section{Transition diagram}

\begin{figure}[ht]
 \centering
  \includegraphics[width=.35\textwidth]{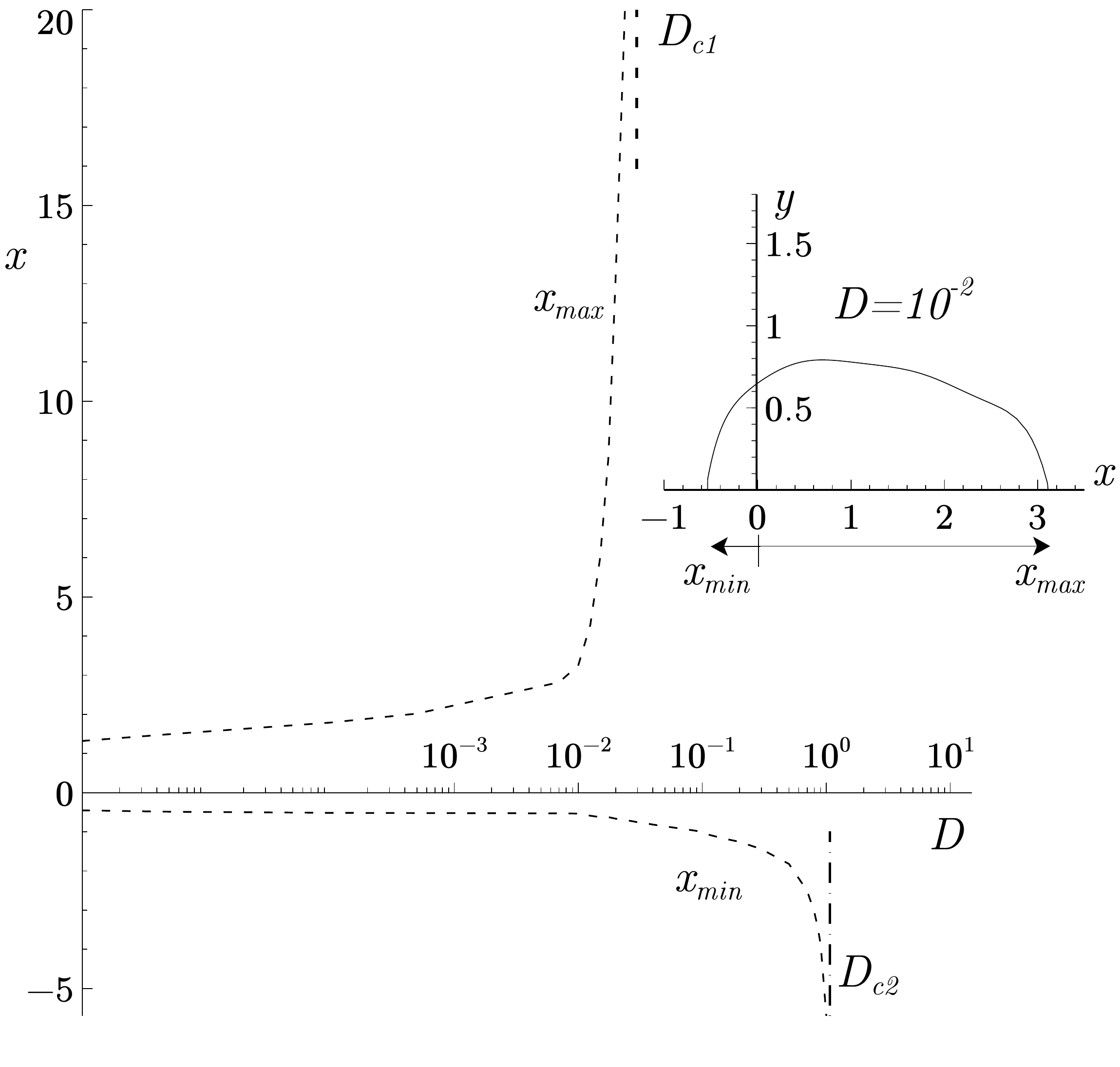}
  \caption{The variation with $D$ of the reactive-kernel boundaries $x_{max}$ and $x_{min}$ for $Le=1$ and $q=0.1$; the inset displays the reaction contour $r_{max}(\theta)$ for $D=0.01$.}
\label{fig:transition_example}
\end{figure}

The plot of $r_{max}(\theta,t)$ for $t \gg 1$ provides for $D < D_{c1}$ the closed boundary $r_{max}(\theta)$ of the steady reactive kernel, with $x_{max}=r_{max}(0)$ and $x_{min}=-r_{min}(\pi)$ defining the downstream and upstream boundaries of the kernel, respectively. The variation of these two quantities as a function of $D$ is shown in Fig.~\ref{fig:transition_example} for $q=0.1$, and $Le=1$, with a sample computation of the complete kernel boundary for $D=0.01$ shown in the inset. 

As can be seen, the kernel length $x_{max}-x_{min}$ increases for increasing $D$. As $D$ approaches $D_{c1}$ the downstream boundary $x_{max}$ diverges, as corresponds to the transition from a confined reactive kernel to an anchored deflagration. Similarly, the upstream boundary is seen to diverge as $D$ approaches $D_{c2}$, the steady solution being replaced by an upstream propagating front.  

\begin{figure}[ht]
 \centering
  \includegraphics[width=.35\textwidth]{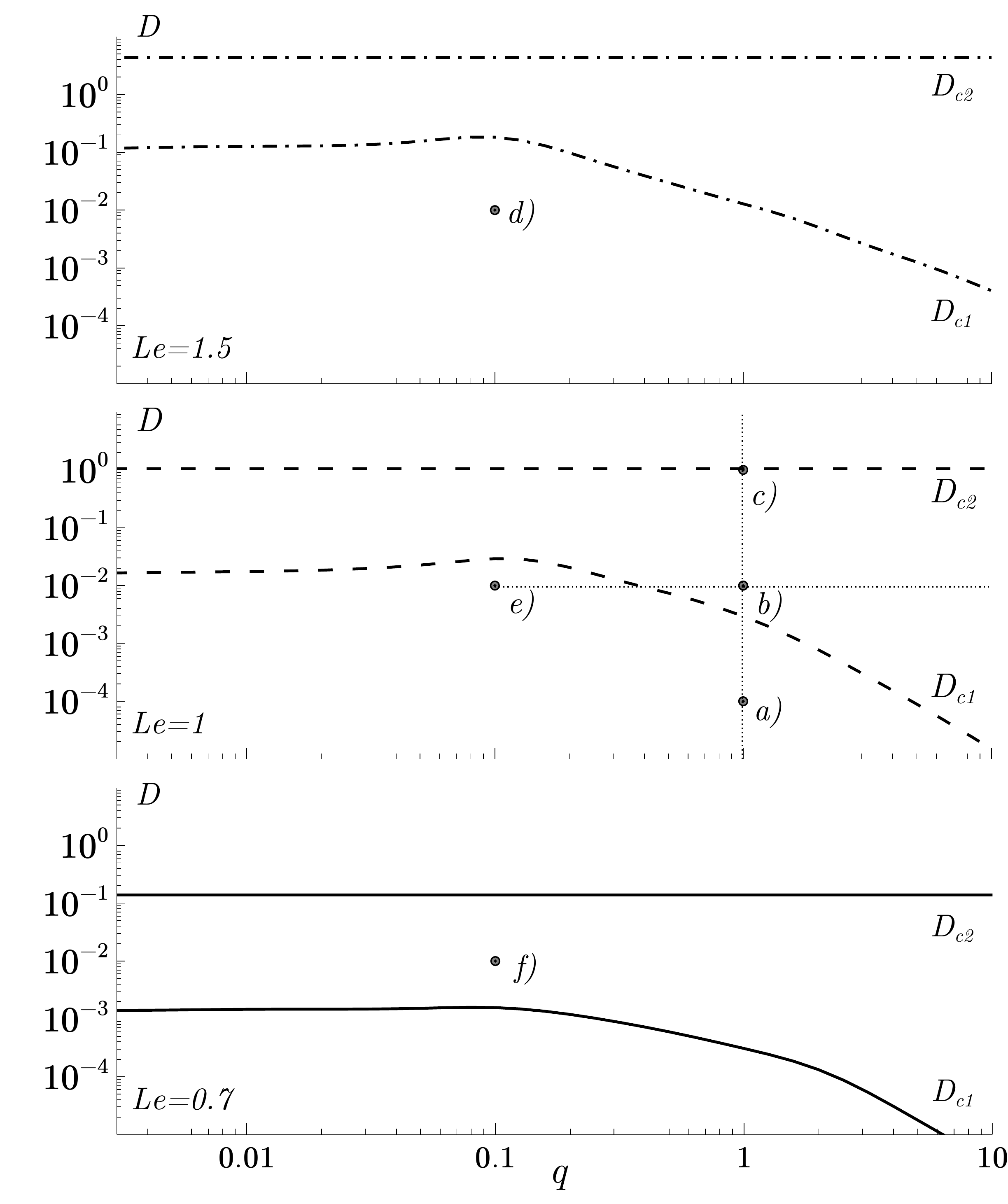}
  \caption{The ignition diagram $D-q$ for different values of $Le$ with indication of the conditions corresponding to the calculations shown in Fig.~\ref{fig:sample}.}
\label{fig:map}
\end{figure}

The critical values $D_{c1}$ and $D_{c2}$ obtained numerically, which define the parametric regions of existence of the different regimes identified above, are shown in Fig.~\ref{fig:map}. The six cases considered in Fig.~\ref{fig:sample}, all corresponding to nonpropagating flames, are indicated in the figure. As previously mentioned, while the value of $D_{c1}$  is a function of both $q$ and $Le$, the value of $D_{c2}$ depends only on the fuel Lewis number, giving for instance $D_{c2}=(4.38,1.07,0.12)$ for the three cases $Le=(1.5,1,0.7)$ considered in Fig.~\ref{fig:map}. This value defines the critical conditions for flashback, a problem considered earlier \cite{KFL_2000,KL_2002}. While the present analysis focuses on the transient initiation process, these previous computations employed a reference frame moving with the front and determined the propagation velocity of the flame relative to the wall as an eigenvalue, with flashback corresponding to the condition of zero propagation velocity\footnote{The values of $D_{c2}$ given here differ slightly from those reported earlier in \cite{KFL_2000,KL_2002} because in the previous computations the Damk\"ohler number was defined as $D=U_L^2/(\alpha_{\scriptscriptstyle T} A)$ in terms of the propagation velocity of the planar flame $U_L$ determined numerically for a finite given value of $\beta$, whereas in the present work we use $D=S_L^2/(\alpha_{\scriptscriptstyle T} A)$ based on the propagation velocity $S_L$ obtained at leading order in the asymptotic solution for large Zeldo'vich numbers. As shown in \cite{Sanchez2014}, for $\beta=10$ the associated correction factor is $(U_L/S_L)=(0.95,0.929,0.891)$ for $Le=(0.7,1,1.5)$, which accounts for the differences observed in the values of $D_{c2}$.}

The comparison of the three panels in Fig.~\ref{fig:map} indicates that the values of $D_{c1}$ and $D_{c2}$ are very sensitive to the fuel diffusivity, both critical parameters increasing by about an order of magnitude as the fuel Lewis number changes from $Le=0.7$ to $Le=1$ and also from $Le=1$ to $Le=1.5$. As expected, for sufficiently large values of $q$ the value of $D_{c1}$, marking the transition from a confined kernel to an anchored flame, decreases with increasing heating rates. The plots also indicate that the value of $D_{c1}$ becomes independent of $q$ for $q \ll 1$, corresponding to weak sources with a heating rate $q'$ much smaller than the heat-release rate associated with the chemical reaction, measured by $\kappa (T_e-T_0)$. In this limiting case the confined reactive kernel becomes a small cylindrical flame ball with negligible effects from external heat addition, and the critical Damk\"ohler number $D_{c1}$ correspondingly becomes independent of $q$. Theoretical descriptions of these flame-ball structures using the limit $\beta \gg 1$ to derive analytical predictions of $D_{c1}$ are worth pursuing in future work.

\section{Conclusions}

The initiation of a deflagration in a premixed boundary-layer stream by continuous heat deposition from a localized energy source has been examined on the basis of one-step Arrhenius chemistry model. The inverse of the boundary-layer velocity gradient at the wall provides the characteristic mechanical time of the problem, which is compared in the controlling Damk\"ohler number $D$ with the relevant chemical time, defined for the reactive mixture from the residence time across the corresponding steady planar deflagration. Three markedly different reaction regimes are identified depending on the value of $D$, namely, (i) for values of $D$ below a first critical value $D_{c1}$, a function of the fuel Lewis number and of the source heating rate, the chemical reaction remains confined in the near-source region at all times; (ii) for values of $D$ larger than  $D_{c1}$ but smaller than a second critical value $D_{c2}$, the latter independent of the heating rate but very sensitive to the fuel diffusivity, the hot kernel around the heat source serves to anchor a deflagration that extends indefinitely downstream; and (iii) for $D>D_{c2}$ the heat deposition generates a deflagration that propagates upstream from the source along near-wall region as a curved flashback front. The effect of the thermal expansion on the velocity field  along with effects of heat losses to the wall, not considered in this preliminary constant-density analysis, are expected to result in order-unity changes of critical Damk\"ohler numbers and will be addressed in future work.

The transition map given above in Fig.~\ref{fig:map} corresponds to laminar-flow conditions, found for instance in the small-scale rotary engines discussed in the introduction, where the Reynolds number based on the thickness of the slender gap that serves as combustion chamber and on the rotor wall velocity is of the order of a few hundred, not high enough to promote transition to turbulence, so that the transient combustion process occurs in these systems in a predominantly laminar environment. It is worth noting that, in systems involving a turbulent boundary-layer flow, heat conduction from the source is confined to the base of the viscous sublayer, so that the above analysis continues to be relevant, provided that $A$ is taken to be the velocity gradient at the base of the viscous sublayer.

\section*{Acknowledgements}

This work was supported by the Spanish MCINN through projects \# CSD2010-00011, ENE2012-33213 and ENE2015-65852-C2-1-R. 

\let\itshape\upshape
\bibliographystyle{ieeetr} 
\bibliography{paperbib_tuneao}
\addcontentsline{toc}{section}{References}

\end{document}